\documentclass{aa}
\usepackage{graphicx}
\usepackage{txfonts}
\usepackage{aalongtable}
\usepackage[round]{natbib}
\begin{document}
   \title{Intracluster Light in the Virgo Cluster: Large Scale Distribution
\footnote{Based on data collected with the 2.5 m Isaac Newton
Telescope on La Palma, the Subaru telescope, operated by the
National Astronomical Observatory of Japan, and the ESO/MPI 2.2 m telescope at
La Silla, Chile, operated by ESO during observing runs 62.N-0248 and
70.B-0086(A)}}

   \author{N. Castro-Rodrigu\'ez\inst{1}, M. Arnaboldi\inst{2,3},
   J. A. L. Aguerri\inst{1}, O. Gerhard\inst{4}, S. Okamura\inst{5},
   N. Yasuda\inst{6} \& K. C. Freeman\inst{7}}

   \offprints{N. Castro-Rodrigu\'ez}

\institute{Instituto de Astrof\'{\i}sica de Canarias, C/O V\'{\i}a
   L\'actea s/n, E-38200 La Laguna, Spain: ncastro@iac.es,
   jalfonso@iac.es 
\and European Southern Observatory, Karl-Schwarzschild-Strasse 2,
D-85748 Garching, Germany: marnabol@eso.org
\and INAF, Osservatorio Astronomico di Pino Torinese, I-10025 Pino
   Torinese, Italy
\and Max-Planck-Institut Institut f\"ur Extraterrestrische Physik,
Giessenbachstrasse, D-85741 Garching, Germany: gerhard@mpe.mpg.de
\and Department of Astronomy and Research Center for the Early
Universe, School of Science, University of Tokyo, Tokyo 113-0033,
Japan: okamura@astron.s.u-tokyo.ac.jp
\and Institute for Cosmic Ray Research, University of Tokyo, Kashiwa,
Chiba 277-8582, Japan: yasuda@icrr.u-tokyo.ac.jp
\and Research School of Astronomy and Astrophysics, Mount Stromlo
Observatory, Cotter Road, Weston Creek, ACT 2611, Australia:
kcf@mso.anu.edu.au}
 
   \date{Received 13.03.2008; Accepted 11.08.2009}

   \authorrunning{N. Castro-Rodrigu\'ez et al.}
   \titlerunning{Intracluster Light in the Virgo Cluster}

% \abstract{}{}{}{}{} 
% 5 {} token are mandatory
 
\abstract 
% context heading ({optional) 
% {} leave it empty if necessary 
{} 
% aims heading (mandatory)  
{The intracluster light
  (ICL) is a faint diffuse stellar component in clusters made of stars
  not bound to individual galaxies. We have carried out a large scale
  study of this component in the nearby Virgo cluster.}  
% methods heading (mandatory)  
{The diffuse light is traced using planetary
  nebulae (PNe). The surveyed areas were observed with a narrow-band
  filter centered on the [OIII]$\lambda 5007$ \AA\ emission line
  redshifted to the Virgo cluster distance (the on-band image), and a
  broad-band filter (the off-band image). For some fields, additional
  narrow band imaging data corresponding to the H$\alpha$ emission
  were also obtained.  The PNe are detected in the on-band image due
  to their strong emission in the [OIII]$\lambda 5007$ \AA\ line, but
  disappear in the off-band image. The contribution of Ly-$\alpha$
  emitters at $z=3.14$ are corrected statistically using blank field
  surveys, when the H$\alpha$ image at the field position is not
  available.}  
% results heading (mandatory) 
{We have surveyed a total area of 3.3 square degrees in the Virgo
  cluster with eleven fields located at different radial distances.
  Those fields located at smaller radii than 80 arcmin from the
  cluster center contain most of the detected diffuse light. In this
  central region of the cluster, the ICL has a surface brightness in
  the range $\mu_{B} \,= \, 28.8 - 30$ mag~arsec$^{-2}$, it is not
  uniformly distributed, and represents about 7\% of the total galaxy
  light in this area.  At distances larger than 80 arcmin the ICL is
  confined to single fields and individual sub-structures,
  e.g. in the sub-clump B, the M60/M59 group. For several fields at 2 and 3
  degrees from the Virgo cluster center we set only upper limits.}
% conclusions heading (optional), leave it empty if necessary 
{These results indicate that the ICL is not homogeneously distributed
in the Virgo core, and it is concentrated in the high density regions
of the Virgo cluster, e.g. the cluster core and other
sub-structures. Outside these regions, the ICL is confined within
areas of $\sim 100$ kpc in size, where tidal effects may be at
work. These observational results link the formation of the ICL with
the formation history of the most luminous cluster galaxies.}


   \maketitle
%
%________________________________________________________________

\section{Introduction}

Galaxy clusters are the most massive gravitationally bound structures
known in the Universe. In a volume of a few Mpc$^{3}$, they contain a
large number of galaxies, which interact with each other, the hot gas
and with the global cluster gravitational potential. Thus, in these
high density regions the evolution of galaxies may be influenced by
the environment.

With new numerical algorithms and fast computers, we are able to
simulate large volumes of the Universe and follow the evolution of
galaxies in different environments as function of redshift. These
simulations show that galaxies undergo important transformations over
time in cluster environments. They can evolve over short timescales
(less than 1 Gyr) due to different mechanisms such as fast
interactions between galaxies and with the cluster gravitational
potential \citep{merritt84,moore96}, gas stripping \citep{gunn72,
quilis00} and starvation \citep{larson80, bekki02}. Moreover, in the
cluster central region, where the galaxy density is high, galaxies can
also undergo important transformations over larger timescales due to
galaxy mergers \citep[][and references therein]{mihos04}.  According
to these results, galaxy evolution is not frozen at the earliest
phases of the cluster formation, as described in earlier works
\citep{richstone83,merritt84}, but is still active today.

Hierarchical structure formation theories predict that
massive galaxy clusters are built through the infall of matter,
i.e. galaxies, groups and sub-clusters, along large scale filaments
\citep{white78, west95}. Since mass accretion is still active today,
nearby galaxy clusters may be at different epochs of their mass
assembly, and they may still be in a un-mixed state. The
dynamical state of a galaxy cluster has been traditionally measured by
determining the degree of substructures in the galaxy distribution and
in the X-ray emitting hot intracluster medium, but the inferred
results do not always agree. \cite[see][and references
therein]{ramella07}

%_____________________________________________________________
%                                             Two column Table 
%_____________________________________________________________
%
\begin{table*}
\caption{Summary of the field positions, filter characteristics and
  exposure times for the narrow-band (on-band) images.}
\label{table:1}      
\centering          
\begin{tabular}{c c c c c c c c c c c c c} 
\hline\hline
& & & \multicolumn{4}{c}{[OIII]} & & \multicolumn{3}{c}{H$\alpha$} \\
 Field & RA(J2000) & DEC (J2000)& $\lambda_{c}$ & FWHM & Exposure & S$_{FWHM}$& & $\lambda_{c}$ & FWHM & Exposure & S$_{FWHM}$\\
       & (hh mm ss)& ($^{o}$ ' '') & ($\AA\ $) & ($\AA\ $) & (s) & (arcsec) & & ($\AA\ $) & ($\AA\ $) & (s) & (arcsec) \\
\hline                    
Core & 12 27 48 & +13 18 46 & 5023 & 80 &24000 &1.20 &&---	  &---&---   &---	\\
FCJ  & 12 30 39 & +12 38 10 & 5027 & 44 &18000 &1.40 &&---	  &---&---   &---	\\
LPC  & 12 25 32 & +12 14 39 & 5027 & 60 &27000 &1.40 &&---	  &---&---   &---	\\
SUBC & 12 25 47 & +12 43 58 & 5021 & 74 &3600  &0.68 && 6607      &101&8728  &0.70   	\\
\hline
LPRX & 12 28 00 & +09 30 00 & 5027 & 60 &27000 &1.80 &&---	  &---&---   &---	\\
LPE  & 12 17 27 & +13 41 30 & 5027 & 60 &27000 &1.30 &&---	  &---&---   &---	\\ 
LPS  & 12 27 34 & +10 43 42 & 5027 & 60 &27000 &1.30 &&---	  &---&---   &---	\\
LSF  & 12 38 33 & +12 08 41 & 5023 & 80 &33000 &1.08 &&---	  &---&---   &---	\\
RCN1 & 12 26 13 & +14 08 03 & 5023 & 80 &24000 &1.20 && ---       &---&---   &---	\\
SUB2 & 12 24 13 & +14 09 01 & 5021 & 74 &3600  &1.00 && 6607      &101&9000  &1.10	\\
SUB3 & 12 24 32 & +15 21 21 & 5021 & 74 &3600  &1.00 && 6607      &101&9000  &1.05	\\
\hline                  
\end{tabular}
\begin{list}{}{}
\item[] The fields listed above the horizontal line are located within
80 arcmin of the Virgo cluster center; the data reduction is
described in previous work (FCJ: Arnaboldi et al. 2002; SUBC:
Arnaboldi et al. 2003; Core and LPC: Aguerri et al. 2005). The fields
listed below the horizontal line are at distances between 80 and 215
arcmin of Virgo cluster center.
\end{list}           
\end{table*}

%-------------------------------------------------------------

\begin{table*}
\caption{Summary of the filter characteristics and exposure times for
  the broad-band (off-band) images in the various fields.}
\label{table:2}      
\centering          
\begin{tabular}{c c c c c c c c c c c c c} 
\hline\hline
&  \multicolumn{4}{c}{B-band}  & \multicolumn{4}{c}{V-band} &
\multicolumn{4}{c}{R-band}\\
 Field & $\lambda_{c}$ & FWHM &Exposure & S$_{FWHM}$ & $\lambda_{c}$ & FWHM &Exposure & S$_{FWHM}$ & $\lambda_{c}$ & FWHM &Exposure & S$_{FWHM}$  \\

  & ($\AA\ $) & ($\AA\ $) & (s) & (arcsec) & ($\AA\ $) & ($\AA\ $) & (s) & (arcsec) & ($\AA\ $) & ($\AA\ $) & (s) & (arcsec) \\ 
\hline                    
Core &---  &---  &---  &--- &  5395 & 894 &2400  & 1.20 &---  &---  &--- &--- \\
FCJ  &---  &---  &---  &--- & 5300 & 267 & 6000 & 1.4 &---  &---  &--- &--- \\
LPC  &  4407 & 1022&5400  & 1.50&---  &---  &--- & --- &--- &--- &--- & ---\\
SUBC &---  &---  &  & & 5500  &956  &900   & 1.00 &6500  &1130  &720 &0.86 \\
\hline
LPRX &  4407 & 1022&15600  & 1.60&---  &---  &--- & --- &--- &--- &--- & ---\\
LPE  &  4407 & 1022&15600 & 1.50&---  &---  &--- & --- &--- &--- &--- & ---\\ 
LPS  &  4407 & 1022&15600 & 1.50&---  &---  &--- & --- &--- &--- &--- & ---\\
LSF  &---  &---  &---  & ---& 5395 & 894 &14300 & 1.10& --- &---  &--- &--- \\
RCN1 &---  &---  &---  & ---& 5395 & 894 &2400  & 1.20& --- &---  &--- & ---\\
SUB2 &---  &---  &---  & ---&  5500 & 956 &3600  & 1.10&6500  &1130  &1440 &0.80 \\
SUB3 &---  &---  &---  & ---&  5500 & 956 &3600  & 0.80& 6500 &1130  &1440 &0.90 \\
\hline       
\end{tabular}
\end{table*}

The gravitational forces acting on galaxies during the cluster
formation and its evolution unbind a fraction of their stars, which
then end-up orbiting in the intracluster region. This unbound diffuse
stellar component is the so-called intracluster light (ICL). Numerical
simulations show that the amount of ICL in clusters depends on the
cluster mass and dynamical state
\citep{murante04,willman04,sommerlarsen05} .  A more massive, older,
dynamically evolved cluster may contain a larger amount of ICL than a
less massive or a dynamically younger system
\citep{rudick06,murante07}. Furthermore, in a highly evolved
dynamically old cluster the ICL morphology would be more diffused, with
relative few streams, while a dynamically young cluster dominated by
groups still in the process of merging is likely to be dominated by
ICL found in streams \citep{rudick09}. Therefore the study of the
amount, distribution and kinematics of the ICL may provide information
on the cluster accretion history and evolutionary state, as well as
about the evolution of cluster galaxies.

The ICL has been mapped using several observational techniques, e.g.
deep broadband imaging \citep{feldmeier02,feldmeier04a, mihos05,
zibetti05, krickber07}, and the detection of individual stars
associated with the diffuse stellar component, both red giant branch
stars \citep{ferguson98,durrell02,williams07} and intracluster
planetary nebulae \citep[ICPNe][]{arnaboldi96,arnaboldi02,arnaboldi03,
feldmeier98,feldmeier03,feldmeier04b,aguerri05,aguerri06}. In particular,
ICPNe are the only component of the ICL whose kinematics can be
measured \citep{fac+00,arnaboldi04,gerhard05,gerhard07,doherty09}.

Several studies investigate the properties of the ICL in the core
of the nearby Virgo cluster.  Expanding on our earlier imaging work in
the Virgo cluster core
\citep{arnaboldi02,arnaboldi03,okamura02,aguerri05}, we have now
completed a survey campaign of the ICL distribution on larger
scales, outside the center of the Virgo cluster.  In total, we have
now covered more than 3 square degrees in Virgo, at eleven different
positions in the cluster and at distances between 80 arcmin and some
100 arcmin from the Virgo cluster center.  In several of the new fields,
the ICL is at least two magnitudes fainter than in the core region.
These new results are in agreement with observations of intermediate
redshift clusters \citep{zibetti05}, and with the results of recent
high resolution hydrodynamical simulations of cluster formation in a
$\Lambda$ CDM universe \citep{murante04,rudick06,murante07}, which
predict that the ICL is more centrally concentrated than cluster
galaxies and that the largest portion of the ICL is formed during the
assembly of the most luminous cluster galaxies.

In the following, we will present the results of a systematic and
homogeneous ICPNe survey in regions outside the Virgo core and discuss
the results jointly with ICPN survey data available for the Virgo
cluster.  In Section~2 we will present the data for the new fields in
the Virgo cluster.  The technique for the selection of ICPNe will be
summarized in Section~3.  The possible contamination of the ICPN
catalog by background emission-line objects is quantified in
Section~4. In Section~5 we use the number of detected ICPNe to infer
the surface brightness of the light at the fields' position and
compare our results with previous works. We discuss the distribution
of the ICL from the Virgo core to the outer distant fields at $\sim
200' = 0.8$ Mpc, and its implications for the current models for the
formation of ICL in clusters, in Section~6. In Section~7 we set limits
on the contribution from a possible homogeneous luminous density in
the nearby Universe to our PN sample, and finally summarize our
conclusions in Section~8.

In what follows we consider the center of the Virgo cluster at the
position of Cluster A \cite{binggeli87}, at the peak of the luminosity
density of all galaxies in the Virgo cluster, e.g. $\alpha(J2000) =
12:27:49.95$\, $\delta(J2000)=13:01:24.57$. We adopt a distance of 15
Mpc to the Virgo cluster center, thus $1'' = 72.7$ pc.

\section{Observations and Data Reduction}

%-------------------------------------------------------------
   \begin{figure*}
   \centering
   \includegraphics[width=10cm]{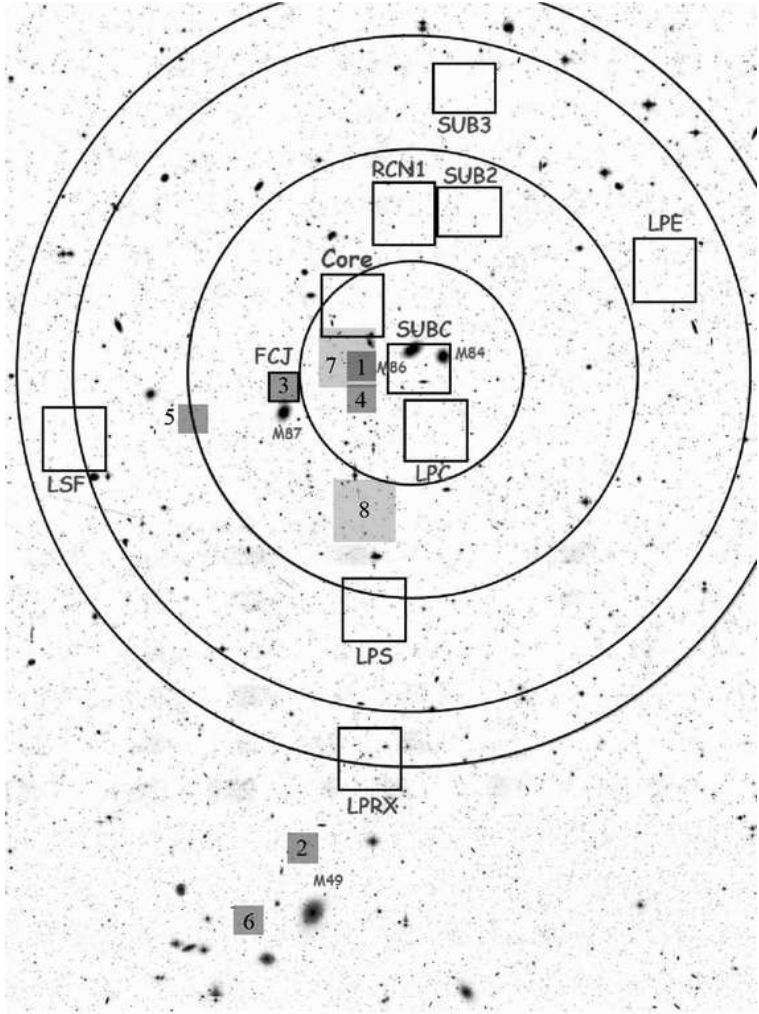}
      \caption{ Finding chart of the Virgo fields discussed in this
paper.  The circles indicate radial distances of 1, 2, 3 and 3.5
degrees from the Virgo cluster center identified by \citet{binggeli87}:
$\alpha(J2000) = 12:27:49.95$\, $\delta(J2000)=13:01:24.57$. Open
squares indicate the field positions whose data were analyzed by our
team. The area covered by each field is : 16$\times$16 arcmin$^{2}$
for the FCJ field, 34$\times$27 arcmin$^{2}$ for SUBC, SUB1 and SUB2
fields. The remaining fields have areas of 34$\times$34 arcmin$^{2}$
(Core, RCN1, LSF and LP* fields). The shaded fields indicate the areas
surveyed by \citet{feldmeier98,feldmeier03,feldmeier04b}; they are designated
as F04$-n$ in the text, where $n$ is the number given in the respective
shaded area on the figure.}\label{mapvirgo}
   \end{figure*}
%
%_____________________________________________________________

Previous ICPN surveys in the Virgo cluster were carried out mostly in
the Virgo core
\citep{feldmeier98,feldmeier03,feldmeier04b,arnaboldi02,
arnaboldi03,aguerri05} and around M49 \citep{feldmeier04b}.

We have now collected additional data in several fields located at
larger radial distances from the center of the Virgo cluster. In total
we have now six new fields, which in addition to those surveyed in
\citet{aguerri05} cover a total area of $3.3$ square degrees. The
Virgo cluster SDSS image and the location of the eleven fields (five
fields surveyed in previous studies by our group and the new six ones)
are shown in Figure~1. We indicate also the areas surveyed by
\citet{feldmeier98,feldmeier03,feldmeier04b} and the name of the
brightest Virgo galaxies.  The results of the survey for ICPNe in the
LPRX, LPE, LPS, LSF, SUB2, SUB3 field are presented here for the first
time. In Table~1 and 2 we give the eleven fields' IDs and locations.

Three of the new fields were observed during two observing runs in
March 2002 and 2004 with the Wide Field Camera (WFC) on the 2.5~m
Isaac Newton Telescope (INT) at Roque de los Muchachos Observatory (La
Palma, see Tables~1 and 2). We refer to them hereafter as the LPRX,
LPE and LPS fields.  The images from this camera consists of a mosaic
of four CCDs with a total field of view of 34$' \times$ 34$'$. The
pixel scale is 0$''$.333, and the mean readout noise and gain of the
detectors are 6.1 ADU pixel$^{-1}$ and 2.8~e$^{-}$~ADU$^{-1}$,
respectively. We imaged these fields through a 60$\AA$ wide filter
centred at 5027$\AA$, which contains the wavelength of the
[OIII]$\lambda5007 \AA$ emission at the redshift of the Virgo
cluster. We will refer to this narrow-band filter as the ''on-band''
filter (see Table~1). We also took an additional broad-band image per
field through a filter centred on the B band (4407$\AA$) and 1022$\AA$
wide. The broad-band filter is referred to as the ''off-band'' (see
Table~2). The total exposures were 27000 sec for the on-band images
and 15600 sec for the off-band.

In April 2004 we acquired another field (LSF) with the Wide Field
Imager (WFI) on the ESO/MPI 2.2 m telescope located at La Silla
Observatory (Chile). The full image from this camera consists of a
mosaic of eight CCD images, covering 34$' \times 34'$ on the sky. Each
CCD has a pixel scale of $0''.238$, and an average readout noise and
gain of 4.5~ADU~pixel$^{-1}$ and 2.2~e$^{-}$~ADU$^{-1}$,
respectively. The field was imaged through an on-band 80$\AA$ wide
filter, centred at 5023$\AA$. In addition we also imaged this field
through an off-band V-band 894$\AA$ wide filter, centred at 5395$\AA$
(see Table~1 and 2).

From January to April 2004 two fields in the Virgo Cluster (SUB2 and
SUB3) were imaged with the Suprime-Cam 10k$\times$8k mosaic camera, at
the prime focus of the 8.2 m Subaru telescope
\citep{miyazaki02}. The field of view of these images covers an
area of 34$'\times$27$'$, with a pixel size of $0''.2$. The fields
were imaged through two on-band filters and two off-band, the
standard V and R filters (see Tables~1 and 2).

The new fields at larger cluster distances from M87 are complementary
to those studied in the Virgo cluster core; their parameters are
reported in Table~1 and 2.  The data reduction and calibration of the
new fields, LPRX, LPE, LPS and LSF, follow the procedure described in
\citet[hereafter Paper I]{arnaboldi02}. The data reduction is
performed using the MSCRED package in IRAF\footnote{For a detailed
discussion of the mosaic data reduction, we refer the reader to
\citet{alcala02}.}.  Calibrations are done using several Landolt
fields for the off-band filters, and spectrophotometric stars for the
on-bands. Fluxes are normalized to the AB magnitude system, following
\citet{theuns97}. A detailed description of the calibrations
steps and the relation between the AB magnitudes and the m(5007)
[OIII] magnitude introduced by \citet{jacoby89} is given in
Paper~I. Table~3 shows the conversion constant (K) between the two
magnitude systems: m(5007)=m$_{AB}$+K.

The data for the additional new fields observed with the SuprimeCam at
the SUBARU telescope (SUB2 and SUB3) are processed using the standard
data reduction package developed for SuprimeCam data. The photometric
calibrations for the SUB2 and SUB3 data are obtained using Landolt
fields for the broad bands, and spectrophotometric stars for the
narrow bands\footnote{Zero points for the Subaru data are
$z_{0,H\alpha} = 24.0$, $z_{0,[OIII]} = 24.6$, and $z_{0,V+R}=27.8$
for a 1 sec exposure}.

For completeness, Table~1 and 2 provide a summary of the fields
positions, filter characteristics and exposure times for on-band and
off band exposures for the surveyed areas in the Virgo core.  Throughout this
work we will use the notation $m_{n}$ and $m_{b}$ to refer the on-band
and off-band magnitudes of the objects in the AB photometric system.

\section{Detection and selection of ICPN candidates}

In this work, the diffuse light in a Virgo cluster field is measured
by the number density of detected PNe. The PNe are identified via
their strong emission in the [OIII]$\lambda 5007 \AA$ line and, in the
two SUB2, SUB3 fields, may be confirmed by the additional detection
of the associated H$\alpha$ emission. The observational techniques
used for the PNe detection have been successfully applied in the Virgo
cluster, in nearby groups \citep[Leo;][]{castrorodriguez03}, and in
compact groups \citep[HCG44;][]{aguerri06}. In a few Virgo fields we
carried out the spectroscopic follow-up, resulting in a large fraction
(up to 80$\%$ for some fields) of spectroscopically confirmed PNe
\citep{arnaboldi04,doherty09}.

%-------------------------------------------------------------
   \begin{figure*}
   \centering
   \includegraphics[width=17cm]{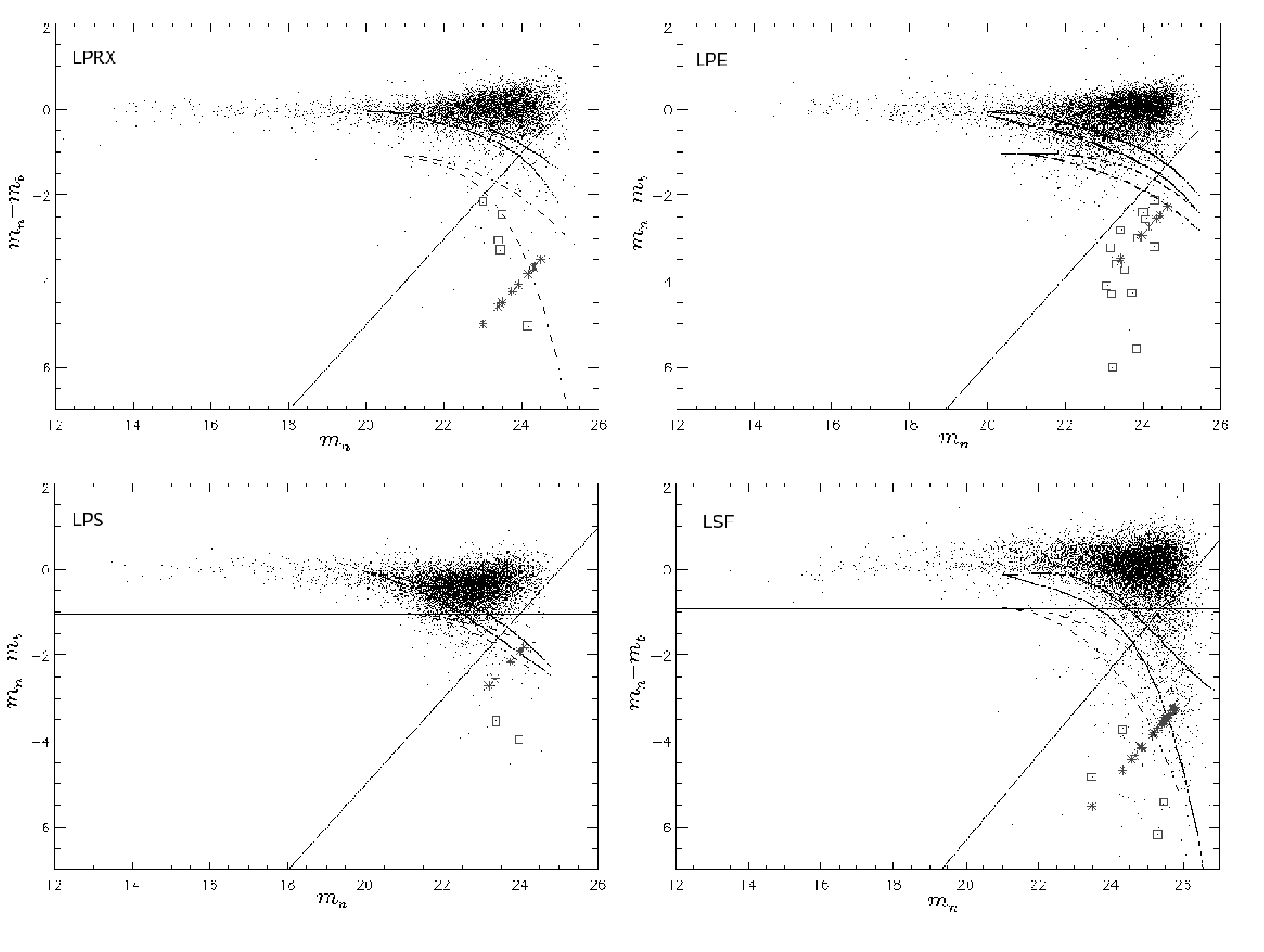}
     \caption{Color-magnitude diagrams (CMDs) for all sources in the
      LPRX, LPE, LPS and LSF fields. The horizontal lines
      indicate the color excess of emission line objects with an
      observed EW=100$\AA$. The inclined continuous lines show the
      $m_b$ magnitude corresponding to 1.0 $\sigma$ above the sky, in
      the off-band images. The curved lines limit the
      regions where 99$\%$ and 99.9$\%$ of the simulated continuum
      objects would fall in this diagram, given the photometric
      errors. The dashed lines limit those regions where 84$\%$ and
      97.5$\%$ of the simulated objects with color excess relative to
      an observed EW=100$\AA$ would fall in this diagram, as function
      of magnitude and the photometric errors.  The asterisks
      represent point-like sources with off-band magnitudes fainter
      than the detection limit in the off-band image as measured by
      SExtractor; the squares represent point-like objects with an
      off-band emission fainter than the limiting magnitude in the
      off-band image.}\label{Fig:2}
   \end{figure*}

\subsection{[OIII] ``on-off'' band survey of ICPNe - Catalog selection with 
the color magnitude diagram (CMD) }

The ICPN candidates in the fields LPRX, LPE, LPS and LSF were
identified using the on--off band technique, pioneered by Ciardullo
et al. (1989) and automatized for large areas by \citet[Paper
I]{arnaboldi02}, using selection criteria based on color excess
discussed by \citet{theuns97}. On the basis of the strong line emission
of a PN at the [OIII]$\lambda5007\AA$ line, the photometric ICPN
candidates can be identified as point-like sources detected in the
[OIII] image, the on-band and, because of their faint continuum, they
cannot be detected in the off-band image. The on-band filter is a few
tens of \AA\ wide filter (60 or 80 \AA, see Table~1 and 2) centred at
the wavelength of the redshifted [OIII]5007\AA\ emission of the PN
population at a given distance.  In Paper I, this technique was
standardized and automatized to select ICPNe in wide field images. We
give a brief summary here.

We detect all objects on the on-band images. We then perform aperture
photometry with SExtractor \cite[][]{bertin96} on the on-band image at
the sources' position, and in the corresponding aperture in the
off-band image. All point-like sources detected on the on-band image
are placed in a color--magnitude diagram (CMD), m$_{n}$-m$_{b}$ vs
m$_{n}$, and are classified according to positions in this
diagram. The ICPN candidates are point-like objects with no detected
continuum emission and observed EW greater than the color excess curve
corresponding to EW=100$\AA$ and convolved with the photometric error,
see Paper I for more details.  Figure~\ref{Fig:2} shows the CMDs for the LPRX,
LPE, LPS, and LSF fields; the selected ICPN candidates are indicated
by asterisk and square symbols.

%_____________________________________________________________
%
\begin{table}
\caption{Limiting magnitudes of the fields in the on-band, off-band
  filter, and the K constant}
\label{table:3}      
\centering          
\begin{tabular}{c c c c}
\hline\hline
Field & m$_{lim}(5007)$ & m$_{b\,lim}$ & K \\
\hline                    
Core &  27.21 & 24.75 & 2.51 \\
FCJ  &  27.01 & 24.58 & 2.50 \\
LPC  &  27.52 & 25.40 & 3.02 \\
SUBC &  28.10 &  ---  & 2.49\\
\hline
LPRX &  27.49 & 25.05 & 3.02\\
LPE  &  27.83 & 25.93 & 3.02\\ 
LPS  &  27.69 & 25.91 & 3.02\\
LSF  &  28.20 & 26.32 & 2.51\\
RCN1 &  26.71 & 25.63 & 2.51 \\
SUB2 &  27.99 & 26.81 & 2.49\\
SUB3 &  27.74 & 26.65 & 2.49\\
\hline                  
\end{tabular}
\end{table}

The magnitude limit of our emission-line catalog is evaluated using
Monte Carlo simulations. We distribute a large sample of point-like
objects ($\sim$ 10\,000) randomly on the scientific images. We then
perform the photometry of the objects using the same procedure as for
the real emission sources. The limiting magnitude is then defined as
the magnitude at which 50$\%$ of the simulated population is
retrieved. In Table~3 we list the limiting magnitudes for the on-band
image in the m(5007) photometric system, and for the off-band
magnitude in the Johnson system. We list also the conversion constant
K between the AB-system and m(5007). According to Table~3 the
``shallowest'' field is RCN1 ($m_{lim}(5007) = 26.71$) and the deepest
field is LSF ($m_{lim}(5007) = 28.20$), with $m_{lim}(5007)$ fainter
than $27.0$ for all the other survey fields.

At the distance of $15$ Mpc, the bright cut-off of the PN luminosity
function (PNLF) falls at the apparent magnitude of
m(5007)$=26.38$\footnote{This is computed using the absolute magnitude
for the PNLF bright cut-off of $M^{*}_{5007}=-4.5$
\citep{ciardullo02b} and a distance modulus of 30.88.} which is
brighter than the limiting magnitude in all surveyed fields.

On the basis of the CMD selection, we detect 9, 20, 7 and 23
emission-line point-like objects brighter than the m$_{lim}(5007)$ in
LPRX, LPE, LPS, and LSF, respectively. These emission sources include
ICPN candidates and background galaxies. In Appendix~A we give the
coordinates and m(5007) magnitudes of these sources.

\subsection{[OIII] and H$\alpha$ survey of ICPNe - Catalog selection with 
the 2-color diagram (2-CD)}

A more secure selection of the ICPN candidates is possible in the in
the SUB2 and SUB3 fields, using the additional information from the
$H{\alpha}$ images which eliminates the contamination from background
galaxies \citep{okamura02,mihos09}. The selection procedure is
described in \citet[hereafter Paper II]{arnaboldi03}, on the basis of
the instrumental magnitudes measured for sources in the fields,
normalized to 1~sec exposure.

\begin{figure}
   \centering \includegraphics[width=8cm]{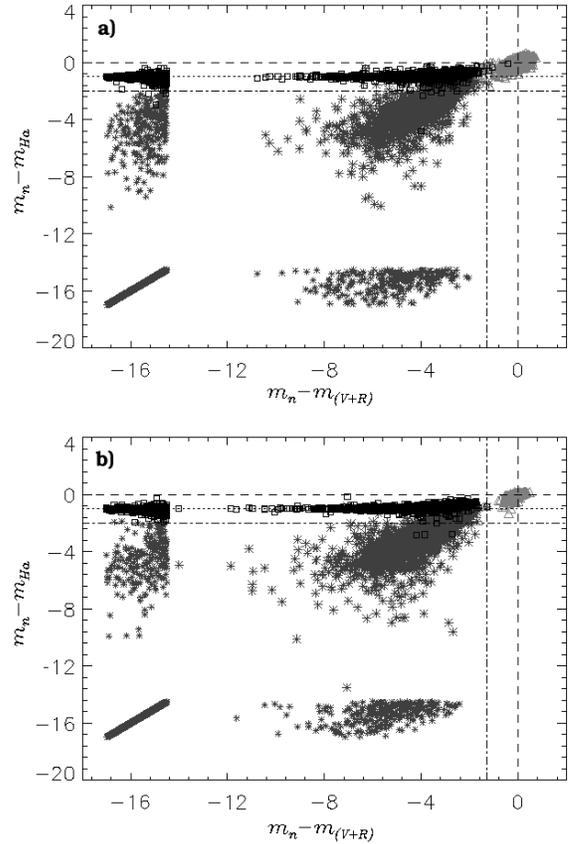}

      \caption{Color--color diagrams ($m_n -m_{H\alpha}$) vs.( $m_n
      -m_{(V+R)}$) of simulated objects for SUB2 (a) and SUB3 (b). The
      different symbols represent: continuum objects (open triangles),
      single line emitters (asterisks), and PNe (open
      squares). Vertical and horizontal dashed lines show objects with
      $m_n - m_{H\alpha} = m_n - m_{(V+R)}=0$.  The vertical
      dashed-dotted line represents objects with $m_n -
      m_{(V+R)}=1.3$. The horizontal dotted and dash-dotted lines show
      objects with $m_n -m_{H\alpha}=-1$ and $-2$ for SUB2 and
      SUB3. The continuum and/or H$\alpha$ magnitudes of the
      simulated objects with no continuum and/or no H$\alpha$ detection
      were set arbitrarily to 40.\label{fig3}} 
   \end{figure}

We use SExtractor for the detection of point-like objects in SUB2 and
SUB3 [OIII] on band images.  Once the objects are extracted, we
measure fluxes at the sources' positions in the H$\alpha$, V and R
band images in matching apertures. We then use a two-color diagram
(2-CD), [OIII]-H$\alpha$ vs. [OIII]-(V+R), for the selection of the
ICPN candidates. In this 2-CD, continuum sources, single-line and
two-line emitters, either PNe or compact HII regions
\citep{gerhard02}, occupy different regions of the plane. We establish
some ad-hoc selection rules for this specific set of images, and the
detection/photometry described above. Simulated population of PNe,
single line emitters, and continuum sources are constructed in the
2-CD, and are used to outline the regions inhabited by the different
kind of objects.  Differently from Paper~II, we have the zero points of
the photometric calibration for the narrow bands, [OIII] and
H$\alpha$, and for the broad band images, V and R, and the selection
criteria for the different classes of objects translate into these
color constraints:

\begin{itemize}
\item The simulated PN population consists of point-like objects with
 a color excess in [OIII], e.g. $m_n-m_{(V+R)}<-1.3$, and have line
 ratios $[OIII]/(H\alpha+[NII])$ larger than $3$
 \citep{ciardullo02b}, because they populate the brighter part of the
 PNLF.  This line ratio translates into an [OIII] vs. H$\alpha$ color
 excess of $m_n-m_{H\alpha} <-1$. Therefore the ICPN-like candidates
 are  point-like sources detected in both [OIII] and H$\alpha$ images,
 whose colors are $m_n-m_{(V+R)}<-1.3$ and $m_n-m_{H\alpha} <-1$; the
 simulated objects are plotted as open squares in Fig.~\ref{fig3}.  .
\item The continuum sources are simulated with a flat spectral energy
  distribution which translates into $m_n-m_{H\alpha}=m_n-m_{(V+R)}=0$
  colors; these are plotted as open triangles in Fig.~\ref{fig3}.
\item Single line emitters in [OIII] are those objects detected in the
  [OIII] image, and have upper limits in either H$\alpha$ or
  in the V+R continuum image, or have no detections in those
  images. These are plotted as asterisks in Fig.~\ref{fig3}. Bright
  $m_n$ sources without H$\alpha$ emission are background, large EW
  Ly-$\alpha$ emitters, while fainter $m_n$ sources without H$\alpha$
  emission can be either PNe or background objects. Because of the
  different detection limits and noise in the H$\alpha$ and V+R image,
  they are scattered in the different regions of the 2-CD. See
  Paper~II for additional details.
\item Compact HII regions are sources with faint continuum,
  $m_n-m_{H\alpha}>0$ and $m_n-m_{(V+R)}<0$, because of the stronger
  H$\alpha$ line emission. 
\end{itemize}

The different class of simulated objects are added to the scientific
images and then recovered using the same automatic procedure as for
the real objects. Fig.\ref{fig3} shows the 2-CDs of the simulated objects
for the SUB2 and SUB3 fields. 

%-------------------------------------------------------------
%
%_____________________________________________________________

The limiting magnitude of the [OIII], H$\alpha$, V and R band images
are listed in Table~3. The $m_{lim}(5007)$ for the on-band [OIII]
images of the SUB2 and SUB3 fields is 1.5 magnitude fainter than the
apparent magnitude of the PNLF bright cut-off at 15 Mpc distance.  The
2-CDs and CMDs of the emission objects in SUB2 and SUB3 images are shown in
Fig.~\ref{fig4}. 

\subsection{Comparison between the two selection procedures}

%-------------------------------------------------------------
   \begin{figure*}
   \centering
   \includegraphics[width=17cm]{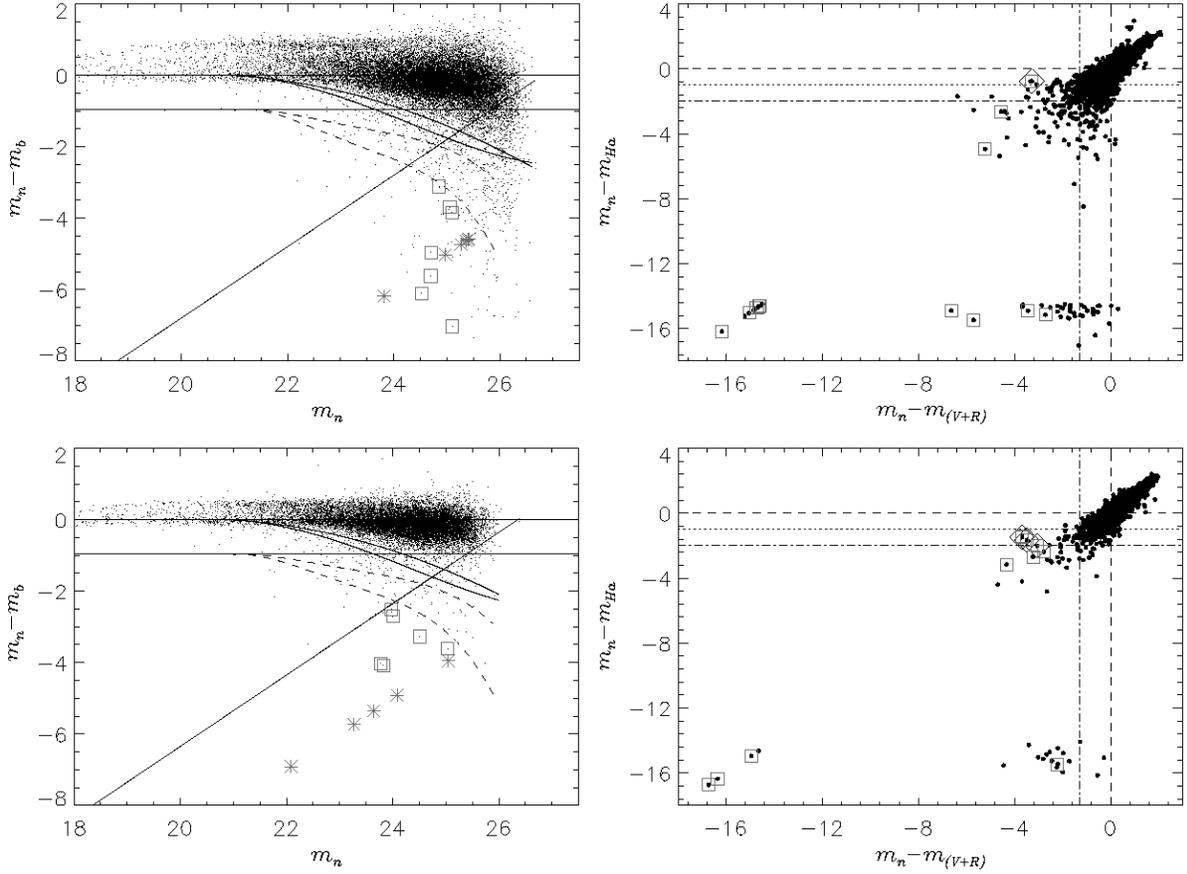}

      \caption{CMD (left panels) and 2-CD (right panels) of the
      sources in the SUB2 field (top panels) and SUB3 field (bottom
      panels) images. On the left plots: the squares and asterisks
      represent objects classified as ICPNe with CMDs. On the right:
      all CMD-identified ICPNe are indicated with open squares; open
      diamonds show the ICPNe selected from the 2-CD. The continuum
      and/or H$\alpha$ magnitudes of those objects with no-continuum
      and/or H$\alpha$ detections were assigned arbitrarily, as in
      Fig.~\ref{fig3}.\label{fig4}}
   \end{figure*}
%
%_____________________________________________________________

For the SUB2/SUB3 fields, we can compare the two methods for
detecting ICPNe either with CMD and 2-CDs. Fig.~\ref{fig4} shows the
CMD for these two fields: we have obtained 13 and 11 objects in SUB2
and SUB3 brighter than the $m_{lim}(5007)$ in these fields, that are
classified as ICPNe on the basis of a color excess criteria in the
CMD. Fig.~\ref{fig4} also shows the distribution of those objects on
the 2-CDs. It turns out that most of the objects classified as ICPNe
in the CMDs are located in the region occupied by single line emitter
in the 2-CDs, i.e. they do have a detectable H$\alpha$ emission at the
position of the [OIII] source on the H$\alpha$ image.  Four objects,
one in SUB2 and three in SUB3, which are ICPN candidates according to
the CMDs, are also two-line emitters according to the 2-CD.

From the 2-CDs, we conclude that we detect one ICPN in SUB2 and three
ICPNe in SUB3 brighter than the limiting magnitudes in [OIII] and
H$\alpha$; the four ICPNe are indicated by open diamonds in the right
panel of Fig.~\ref{fig4}. Furthermore, no HII compact regions are
detected in these two fields.

In empty fields, the sources selected on the basis of a color excess
in a narrow band, e.g. in the $4500 - 5000$ \AA\ wavelength interval,
vs. a broad band filter selected catalog, are most probably
Ly-$\alpha$ background galaxies at redshift $\simeq 3.1$. The
identification of these sources as Ly-$\alpha$ emitters has been
confirmed with spectroscopic follow-up
\citep{kudritzki00,castrorodriguez03,gawiser07,rauch08}.  Therefore
when only a single deep [OIII] image is available to produce the
catalog of point-like emission line objects for a given Virgo field,
we need to subtract statistically the background population of the
Ly-$\alpha$ emitters to select the over-density of objects associated
with the PN population.

We now discuss the statistical subtraction of such background
population in the Virgo cluster fields in Section~\ref{background}.

%-------------------------------------------------------------
   \begin{figure*}
   \centering \includegraphics[width=17cm]{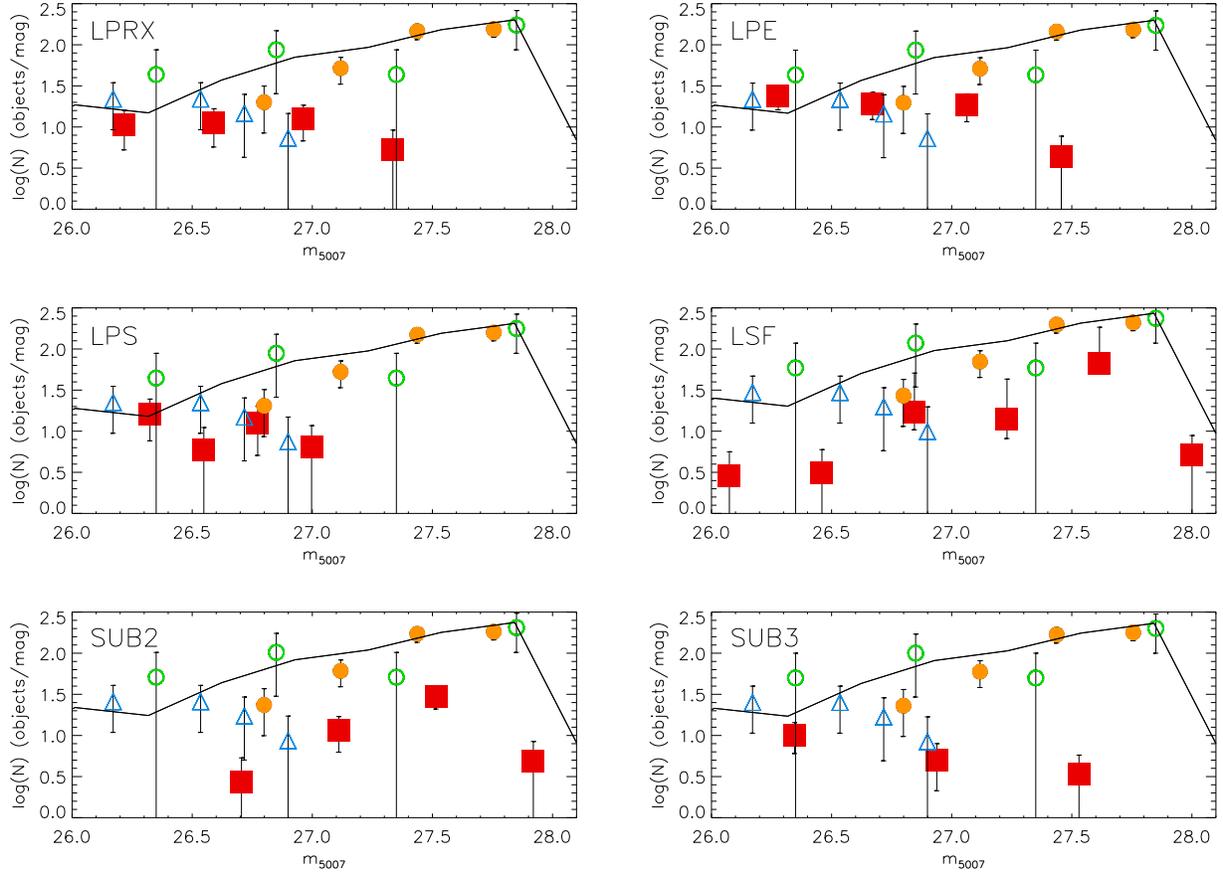}

      \caption{The luminosity function (LF) of the emission-line
      objects detected in LPRX, LPE, LPS, LSF, SUB2 and SUB3 is
      indicated by large full red squares. For comparison, the
      Ly-$\alpha$ LF from \citet{kudritzki00} (open green circles),
      \citet{ciardullo02a} (open blue triangles) and
      \citet{castrorodriguez03} (the Leo field; filled orange circles)
      are also reported in each panel. The continuum line shows the
      luminosity function of the Ly-$\alpha$ candidates from
      \citet{gronwall07}. The LFs of the Ly-$\alpha$ emitters were
      scaled to the effective surveyed area and wavelength range of
      the different observed fields.\label{lfalpha}}
      
         %\label{}
   \end{figure*}
%
%____________________________________________________________

\section{On the nature of the emission-line sources in the Virgo outer 
fields\label{background}}

In addition to a population of ICPNe at the Virgo cluster distance, in
the CMD-selected source list with color excess, we expect a
contribution from misclassified faint continuum objects, [OII]
emitters at intermediate redshift and Ly-$\alpha$ background galaxies
at z$\approx 3.1$. We discuss the contribution from faint continuum
sources in Section~\ref{spillover}, and background galaxies in
Section~\ref{contbck} with the goal of selecting the most likely PN
population at these field positions.

\subsection{Contamination by faint continuum objects\label{spillover}}

It is possible that our catalogs contain misclassified faint continuum
objects.  Because of the photometric errors and a steeply rising
luminosity function, a non-negligible number of faint continuum
objects, which would have $m_n$ fainter than the limiting magnitude in
the on-band image, may end up having a measured $m_n$ brighter than 
$m_{lim}(5007)$. As the same time, due to the photometric errors, the
measured off-band magnitude may be fainter than what is to be
expected, resulting in a color excess for these continuum sources as
for the ICPNe. Given the large number of stars in the field, the
number of such sources can be significant, if the off-band is not
sufficiently deep.

We named this as the ''spillover'' effect in Paper~III, and we refer
to it for additional details. The number of continuum objects that
``spilled over'' can be estimated from simulations of point-like
continuum objects with a LF extrapolated to 3 magnitudes fainter than
$m_{lim}(5007)$ and distributed randomly on the scientific frames. We
then detect the simulated objects with the same criteria as for the
real ones, and determine the number of simulated objects brighter than
$m_{lim}(5007)$ in regions of the CMD where the ICPN photometric
candidates are located.  After scaling the simulations to the number
of observed continuum objects, the final numbers of spilled-over
contaminants brighter than $m_{lim}(5007)$ of the image are found to
be zero in all fields. This result comes about because of the very
deep off-band images acquired for these fields (see the discussion in
Paper~III).

\subsection{Contribution from background galaxies\label{contbck}}

Background emission-line galaxies are detected in [OIII] on-band
photometric surveys
\citep{ciardullo02a,castrorodriguez03,gronwall07}. According to the
central wavelength and limiting flux of the on-band images, the
surveys are sensitive to the
[OII]~3727\AA\ emission line of star-burst galaxies at z=0.35 and the
Ly-${\alpha}$ line of young galaxies at z=3.1.

{\it [OII] emission line at $z\sim0.35$:} We do not expect a
significant contamination by [OII] emitters at medium redshifts in the
final source catalogs because the emission-line objects have a color
excess which implies observed EWs greater than 100$\AA$. Previous
studies of [OII] emitters at z=0.35 have not found objects with
observed EWs greater than 95$\AA$ \citep{colless90, hammer97,
hogg98}. The long-slit search for low surface brightness Ly-$\alpha$
with fluxes of a few$\times 10^{-18}$ erg s$^{-1}$ cm$^{-2}$ by
\citet{rauch08} also reports a small number density of [OII] emitters
at these faint fluxes.

{\it Ly-${\alpha}$ line emitters at z=3.1:} On-band/off-band
photometric surveys of blank fields provide a non-negligible density
of emission objects \citep{ciardullo02a,gronwall07}. The spectroscopic
follow-up of these emission line objects showed that they are
Ly-$\alpha$ emitters at z=3.1 \citep{gawiser07}.

In what follows we adopt a statistical approach to correct for the
contribution of the z=3.1 Ly-$\alpha$ background objects, by comparing
the number density of the ``PN-like'' objects in the outer Virgo
fields with those measured in ``blank-fields''
\citep{kudritzki00,ciardullo02a,castrorodriguez03,gronwall07},
normalized by the effective sampled volume.

\subsection{Are ICPNe detected in the Virgo fields outside the cluster core?
\label{lyalpha}}

We thus estimate the fraction of background galaxies present in the
current catalogs of point-like sources by comparing their LF and
number densities with those of Ly-$\alpha$ emitters detected in
similar [OIII] on band surveys for objects brighter than
m(5007)$=27.0$ and normalized to the effective sampled volumes.  We
shall compare the different sample for magnitudes brighter than
m(5007)$=27.0$; at fainter magnitudes, the automatic classification of
pont-like(unresolved) vs. extended(resolved) based on total magnitude
vs. small central aperture magnitude plot (see Paper I) becomes
unreliable and not uniform among the fields' ICPN catalogs and the
comparison blank field catalogs.

Fig.~\ref{lfalpha} compares the LFs of the point-like emission-line
objects detected in the LPRX, LPE, LPS, LSF, SUB2 and SUB3 fields with
the LF of the Ly-$\alpha$ population from different surveys. 
We then evaluate the number of expected Ly-$\alpha$ emitters in our
field brighter than m(5007)$=27.0$ from the
\citet{kudritzki00,ciardullo02a,castrorodriguez03} surveys, normalized
to the same sampled volume.  Thus, the expected number of Ly-$\alpha$
galaxies are 65/15/6 in LPRX, 65/15/6 in LPE, 66/16/6 in LPS and
86/20/8 in LSF field.

The blank field survey of \citet{castrorodriguez03} selected
point-like emission line sources with same criteria as the current
sample, and it provides a lower inferred number density of Ly-$\alpha$
than either the \citet{ciardullo02a} or \citet{kudritzki00} surveys.
By statistically subtracting the contribution from the background
population estimated from \citet{castrorodriguez03} to the outer Virgo
fields, we derive an upper limit to the number of ICPNe. The number of
emission-line objects detected in the different fields, and the
expected number of Ly-${\alpha}$ objects from the
\citet{castrorodriguez03} survey brighter than m(5007)$=27.0$ are 6/6,
14/6, 6/6, 3/8 for LPRX, LPE, LPS, and LSF fields, respectively. The
LPE field shows an excess of 8$\pm 4$ emission-line objects over the
expected number of background galaxies in this field, while the LSF
field shows a lower density than expected, by 5$\pm 2$.  We may
therefore conclude that the emission-line objects detected in the
LPRX, LPS, and LSF, fields are compatible with being all Ly-${\alpha}$
background galaxies. The emission-line objects detected in LPE are
compatible with containing a few ICPNe at the $2\sigma$ level, as in
the SUB2/3 fields.

An over-density of emission-line unresolved sources is detected at the
RCN1 field position also; here we select $28$ candidates using the
color excess criteria in the CMD, using a deeper off-band image than
in Paper~I, and we expect a population of $8$ Ly-$\alpha$ emitters
from \citet{castrorodriguez03} scaled number counts. 

On the basis of the previous results, the properties of the unresolved
emission line objects in these outer fields are those of the
Ly-$\alpha$ population at $z=3.14$.  Within our sample, we find that
the flux of the brightest emission sources may vary by a factor 2 in
different fields, i.e. from 26.0 in LSF field to 26.7 in SUB2 field.
The \citet{kudritzki00},\citet{ciardullo02a},
\citet{castrorodriguez03} and \citet{gronwall07} samples show similar
field-to-field variation of the Ly-$\alpha$ LF bright end. Since these
samples come from surveys that may adopt different selection criteria,
we believe that our current results on the field-to-field variations
of the Ly-$\alpha$ LF in the six Virgo fields is more robust because it
is based on homogeneously selected samples of point-like objects, with
deep off-band control images. The variation of the Ly-$\alpha$
properties may be related to the cosmic variance: recent observations
by \citet{ouchi03} show that Ly-$\alpha$ galaxies are already
clustered at z=4.86, which implies even greater clustering at
z$\approx 3$.

\section{ICPN number density and ICL surface brightness in
the Virgo cluster \label{ICPN&ICL}}

On the basis of the statistical subtraction of the contribution from
Ly-$\alpha$ emitters at $z=3.14$, we can conclude that:
\begin{itemize}
\item In the SUB2 and SUB3 fields, we detected 4 ICPNe in both [OIII]
and H$\alpha$. These are all brighter than m(5007)$ = 27.0$,
i.e. within 0.5 magnitudes of the PNLF bright cut-off at the distance
of the Virgo cluster;
\item the number densities and LFs of the emission-line objects
detected in the outer Virgo fields -- LPRX, LPS, and LSF -- are
statistically consistent with those of the population of Ly-${\alpha}$
background galaxies;
\item we detect over-densities in LPE (at the $2\sigma$ level) and
in RCN1 (larger than $3 \sigma$) Virgo fields;
\end{itemize}

On the basis of the number of emission line objects detected in the
LPRX, LPS, LSF fields, we can set upper limits to the ICPNe population
in each field. We statistically subtract the most conservative number
of Ly-$\alpha$ emitters expected in each field based on the
\citet{castrorodriguez03} survey (see discussion in
Section~\ref{lyalpha}), estimate the error-bars from Poisson
statistics, and compute the $1 \sigma$ upper limit for the number of
ICPNe independently for each field.  For LPRX, LPS, and LSF, the
expected number of ICPNe brighter than m(5007)$=27.0$ are
$0^{+2.4}_{-2.4}$, $0^{+2.4}_{-2.4}$, and $-5^{+2.4}_{-2.4}$; for LPE
and RCN1 the number of ICPNe is $8^{+2.8}_{-2.8}$ and $20^{+4}_{-4}$
respectively.

\subsection{From ICPN number density to ICL surface brightness\label{alphaf}}
Since PNe closely follow the distribution of starlight in galaxies
\citep{coccato09}, the bolometric luminosity of the parent stars of
the PN population can be obtained as $L_{bol}=n_{PN}/\alpha$, where
$n_{PN}$ is the number of PNs in each field and $\alpha$ is the
luminosity-specific PN density.  We follow the procedure detailed
below:
\begin{itemize}
\item the $n_{PN}$ for a given field is scaled by a factor:
\begin{equation}
\Delta=\frac{\int^{M^{*}+1}_{M^{*}} PNLF(m) dm}{\int^{m_{lim}}_{m^{*}}
PNLF(m) dm}
\end{equation}
where PNLF(m) is the analytic expression for the PNLF
\citep{ciardullo89}, $M^{*}$ and $m^{*}$ denote the absolute and
apparent magnitude of its bright cutoff, respectively, and $m_{lim}$
is the m(5007) limiting magnitude in each field. This scaling
ensures that we account for all PNs within 1 mag of $M^{*}$.  We then
use the corresponding value $\alpha_{1.0}$ to infer the amount of
bolometric luminosity in our fields.
\item \citet{doherty09} computed the $\alpha_{2.5}$ for the halo of
M87 and the ICL directly. The value of $\alpha_{2.5,M87}$ is $3.1
\times 10^{-9}$ PN L$_\odot^{-1}$; for the ICL $\alpha_{2.5,ICL} = 7.2 \times
10^{-9}$ PN L$_\odot^{-1}$; we refer to \citet{doherty09} for
further details on their computation and uncertainties. 
\item From the analytical expression of the PNLF, we can scale the
  value of the luminosity-specific PN density at 2.5 mag fainter then
  the PNLF bright cut off to 1 mag as $\alpha_{1.0} = 0.25\times
  \alpha_{2.5}$.
\item {\it $\alpha$ parameter for the FCJ field -} The results of the
spectroscopic follow-up by \citet{arnaboldi04} and \citet{doherty09}
show that the halo of M87 is extended out to about 35 arcmin from the
galaxy center. Within that radius, the M87 halo contributes 70\% of
the light as sampled by the number of PNe detected at the position of
the FCJ field.  For this field position, we use a value of
$\alpha_{1.0}= 1.1 \times 10^{-9}$ PN L$_\odot^{-1}$ which is the
average value of $\alpha_{1.0,M87}$ and $\alpha_{1.0,ICL}$, weighted
by the fractions of PNs that are bound to M87 ( 70\%) and those
unbound (30\%). Because the FCJ field is at a distance of about
14~arcmin from the center of M87, we can check the derived surface
brightness value at this field position with the V band surface
brightness measurements of \citet{kormendy08}.  
\item {\it $\alpha$ parameter for the SUBC field -}  we use
$\alpha_{1.0}= 1.3 \times 10^{-9}$ PN L$_\odot^{-1}$, which is the
average of $\alpha_{1.0,M87}$ and $\alpha_{1.0,ICL}$. Here we assume
that M84 has the same $\alpha_{1.0}$ value as M87 and the fractions of
bound and unbound PNs in this photometric sample is 50\% each.
\item {\it $\alpha$ parameter for the outer fields -} For the fields
at larger distances, we use $\alpha_{1.0,ICL} = 4.0 \times 10^{-9}$ PN
L$_\odot^{-1}$.
\item Errors on the inferred luminosities and surface brightness are
estimated from Poisson statistics.
\end{itemize}
We compute the bolometric luminosities at each field position from the
number of observed ICPNe, or their upper limit, and the $\alpha$
parameter.  Using the bolometric correction to the B band
\citep{buzzoni06} and (B-V) colors, we derive the B band luminosities
and surface brightnesses. The adopted B-V colors are $0.9$ for the
fields at large distances, and (B-V) = 1.02 for those fields near
either M87 and M84.  In Table~\ref{table:4}, for each field we list
the number of PNe within $m^*+1$, $N_{PN}$, the resulting mean values
of the B-band luminosity, $\Sigma$, and surface brightness $\mu_B$.

%_____________________________________________________________
%
\begin{table}
\caption{$N_{PN}$ within $m*+1$, B band luminosity density and surface
brightness values in the Virgo fields. In LPC, LPRX, LPS and LSF we
list the upper limits.}
\label{table:4}      
\centering          
\begin{tabular}{c c c c}
\hline\hline
Field & $N_{PN}$  & $\Sigma$ & $\mu_{B}$  \\
      & &10$^{6}$ L$_{B,\odot}$ arcmin$^{-2}$ & mag arcsec$^{-2}$ \\
\hline                    
Core & 17.0    &3.1$\pm$0.7  & 29.0$\pm$0.25 \\
FCJ  & 15.0    &12.2$\pm$3.9 & 27.5$\pm$0.25 \\
LPC  & $< 1.0$ & $<0.2$      & $>32.0 $  \\
SUBC & 14.0    &4.0$\pm$0.7  & 28.71$\pm$0.3 \\
\hline
LPRX & $< 2.4$  &$<0.4$         & $>31.1$ \\
LPE  & 8.0      &1.5$\pm$0.5    & 29.8$\pm$ 0.35 \\ 
LPS  & $< 2.4$  &$<0.4$         & $>31.1$ \\
LSF  & $< 2.4$  &$<0.4$         & $>31.1$ \\
RCN1 & 13.0     &2.4$\pm$0.6 & 29.3$\pm$ 0.3 \\
SUB2 & 1.0      &0.2$\pm$0.2 & 31.8$\pm$ 1.0 \\
SUB3 & 3.0      &0.7$\pm$0.3 & 30.6$\pm$ 0.6 \\
\hline                  
\end{tabular}
\end{table}

\subsection{ICL measurements in the Virgo cluster from previous work: the
\citet{feldmeier04b} sample}
Several regions in the Virgo core and around M49 were surveyed by
\citet{feldmeier98, feldmeier03, feldmeier04b} for ICPNe, covering a
total area of 0.89 deg$^{2}$. The positions of these fields are shown
in Fig.~\ref{mapvirgo}, and a summary of their $\mu_V$ measurements is
provided in Table~\ref{table:5}. They find an excess of emission line
sources in all fields, at about 5 times the density of line emitters
in blank fields, but for Field F04-8, where the emission-line object
number density is consistent with the galaxy background population. To
compare these measurements with ours, we take their $\mu_v$ and add a
(B-V) color term determined from the color of the continuum light of
the nearest bright galaxy; the computed values are shown in col. \#4,
in Table~\ref{table:5}.

\begin{table}
\caption{Surface brightness measurements in the \citet{feldmeier04b} fields.}
\label{table:5}      
\centering          
\begin{tabular}{c c c c }
\hline\hline
Field & $\mu_{ICL,V}$     & $\mu_{ICL,B}$ & B-V \\
      & mag arcsec$^{-2}$ & mag arcsec$^{-2}$ & \\
\hline                    
F04-2  & 27.4 & 28.3 & 0.96  \\
F04-3  & 26.5 & 27.5 & 1.02  \\ 
F04-4  & 27.4 & 28.4 & 1.02  \\
F04-5  & 27.3 & 28.3 & 1.02  \\
F04-6  & 28.1 & 29.1 & 0.96  \\
F04-7  & 28.4 & 29.4 & 1.02  \\
F04-8  & -- & -- & --  \\
\hline                  
\end{tabular}
\end{table}

The field with the brightest surface brightness measurement is F04-3
(FCJ in this work; see also Paper I) at 60 kpc from the center of M87.
This field is in fact within the M87 halo, as shown by the deep
photometry of the Virgo core by \citet{mihos05} and the PN kinematics
\citep{arnaboldi04}. The $\mu_B$ measurement for F04-3 agrees with the
independently derived $\mu_B$ for FCJ in Table~\ref{table:4} and with
the surface brightness profile of M87 measured by \citet{kormendy08}
at the FCJ distance from the center of M87. The spectroscopic
follow-up indicates that the M87 halo contributes 70\% of the light at
this field position, and 30\% of the sample is made of free-floating
stars, i.e. the true ICL component. The surface brightness of the ICL
at the position of the FCJ field is then $\mu_B = 28.8$ mag
arcsec$^{-2}$.

\subsection{How well do ICPNe follow light} 
\citet{mihos09} investigated the correlation between the
ICPN distribution and the diffuse light in the Virgo core
\citep{mihos05}. Their results indicate that there is a correspondence
on large scale ($\sim 100$ kpc), which is more robust when the ICPN
catalogs have a lower fraction of background contaminants, as in the
case of the SUBC field (Paper II and \citet{okamura02}), or in the
higher surface brightness regions within the extended galaxy halos. A
possible source of scatter is also a dependence of the
luminosity-specific PN parameter on stellar populations
\citep{ciardullo+05,buzzoni06}.

The procedure adopted in this work maps the over-density of point-like
emitters with respect to a statistically averaged background
population, in fields whose area is about $150\times 150$ kpc$^2$. As
described in Section~\ref{alphaf}, the number of PNe from each
over-density is mapped into a Bolometric luminosity according to the
most appropriate luminosity-specific PN density parameter value for
that field. Our comparison with the \citet{mihos09} results indicate that
the luminosities computed in this work are robust with respect to the
variety of effects that can act to wash out the correlation between PN
number density distribution and light. 

\section{Discussion}

\subsection{The diffuse light in the Virgo cluster}

Several fields outside the Virgo core region contain a population of
emission sources consistent with background Ly-$\alpha$ objects.
Clear over-density of emission line objects of at least a factor 5
with respect to the averaged Ly-$\alpha$ emitters are detected in the
Virgo core region, around M49 \citep{feldmeier03,feldmeier04b}, and in RCN1
and LPE, at a lower $\sigma$ level.

{\it The ICL and the extended halos from bright galaxies- } The
diffuse light observed in the core of a galaxy cluster contains
several luminous stellar components that add up along the
line-of-sight (LOS) to the cluster center: the extended faint halos of
the brightest galaxies and the ICL contribution, defined as the light
coming from stars not bound to individual galaxies. When computing the
ICL fraction in the Virgo core, the surface brightness measurements
must be corrected for the fraction of stars bound to
the extended halos of individual galaxies.  The results from
\citet{arnaboldi04} and \citet{doherty09} on the LOS velocity
distribution of the PNe indicate that the halos of M84, M86 and M87
are extended, out to 150 kpc in the case of M87.

When we select only true ICPNe, we measure a surface brightness for
the ICL of about $\mu_{B} = 28.8$ mag arcsec$^{-2}$ in FCJ and $29.0 -
29.5$ in Core/SUBC. These surface brightness values are similar to
those inferred from the detection of IC RGB stars
\citep{ferguson98,durrell02,williams07}.  

The results from the spectroscopic follow-up imply that the
measurements of the ICL based on a surface brightness threshold,
e.g. $\mu_v > 26.5$, will lead to an overestimate of the ICL, since it
includes the light contributions from extended galaxy halos.

\subsection{On the elongated distribution of the ICL} 
The \citet{feldmeier04b}'s ``smooth'' elongated distribution for the
ICL, of about 4 Mpc along the LOS to the Virgo core, was derived from
the following observational results:
\begin{itemize} 
\item the relative {\it brighter} cut-off of the IC PNLF in sub-clump
A, around M87, vs. sub-clump B, around M49, placing the sub-clump A at
4 Mpc in front of sub-clump B,
\item same number density of ICPNe in these fields.  
\end{itemize}
The evidence for a substantial depth along the LOS is weakened by the
results that some brightening of the PNLF in the M87 halo may
be intrinsic \citep{arnaboldi08}. The presence of an extended halo
around M49, which is to be expected given the results for M87
\citep{doherty09}, may explain the same number density of PNe around
M49 as in M87, because the M49 halo reaches similar surface brightness 
as the M87 one at these fields' relative distances from the galaxy center.

The upper limits in the LPC, LPS and LPRX fields from this work, and
no-PN detection in F04-8 \citep{feldmeier03} show that there is not a
uniform distribution of ICL all across the region from sub-clump A-M87
to subclump B-M49 \citep{binggeli87}.  The ICL drops below the
detection limit in the region of the sky between these two Virgo
sub-clumps. The current results indicate that the ICL is mostly
associated with, and confined to, these high density regions.

\subsection{The ICL surface brightness radial profile }
In Fig.~\ref{figure6} we now plot the surface brightness values from
Table~\ref{table:4} and \ref{table:5} as function of the field
distances from M87, for an easier comparison with \citet{mihos05} and
\citet{kormendy08} photometry. 

{\it Properties of the ICL distribution in the Virgo cluster -} The
ICPN survey in the Virgo cluster indicates that the ICL is not
homogeneously distributed in the core and at larger cluster radii. In
the core, we have an upper limit for the LPC field, and in Field F04-8
\citep{feldmeier03,feldmeier04b}.  Furthermore the deep image of the
central region of the Virgo cluster shows the presence of complex
structures at all scales \citep{mihos05}.  Both results indicate that
the distribution of the diffuse light in the central region is
inhomogeneous, further supporting the evidence that the core is not
relaxed, see \citet{binggeli93}, Paper~III, and \citet{doherty09}.

The comprehensive summary of all surface brightness measurements in
Figure~\ref{figure6} indicates that most of the diffuse light is
detected in fields located in the core, within a distance of about
80 arcmin from M87. Outside this region, the mean surface brightness
decreases sharply, and the ICL is present in isolated pointings,
e.g. RCN1 and LPE. 

Diffuse stellar light is also measured in ``sub-structures'', around
M49 and in the M60/M59 sub-group.  The fields F04-2 and F04-6 from
\citet{feldmeier04b} are situated in the outer regions of M49, at
about 150 kpc from the galaxy center. These fields may contain PNe
from the halo of M49 and the ICL component, which may have formed
within sub-clump B of the Virgo cluster \citep{binggeli87}. Because
the spectroscopic follow-up in not available yet for these PN
candidates, we cannot quantify the fraction of light in the M49 halo
and ICL for these fields.

Using a technique similar to \citet{durrell02} and
\citet{williams07}, \citet{yan08} detected serendipitously an
excess of point-like objects in two ACS fields at 7 and 15 arcmin from
M60. The number density of the excess field population shows a
decreasing gradient with increasing distance from M60. This population
starts at $m_{i775} = 26.2$ and the average surface brightness is 26.6
mag arcsec$^{-2}$ in the I band in the field at 7 arcmin from
M60. These values are very similar to those measured by
\citet{durrell02} observations of the IC RGB in the Virgo
core. \citet{yan08} argue that they detected a stellar population
associated with the M60/M59 pair. This result may indicate that the
M60/M59 sub-group is yet another region of the Virgo cluster where the
diffuse halos and Intra-group-light are being assembled.

The current result on slope of the radial profile of the ICL is
somewhat different from the slow decline in ICPN number density
described by \citet{feldmeier04b}. A steeper radial ICPN density
profile is now derived because the current survey probes areas between
the sub-clump A and sub-clump B in the Virgo cluster, where only upper
limits are set.

\subsection{Fraction of the ICL light vs. galaxy light in the Virgo core}
When we average the measurements of the ICL in the core region defined
as the circular area of 80 arcmin radius\footnote{Because this
circular area includes regions with only upper limits, our average
surface brightness is lower than that measured by \citet{mihos05} in
the region between M87 and M86.} centred on the Virgo cluster center
given by Binggeli et al. (1987), we obtain a measurement of $\mu_B =
29.3 $ mag arcsec$^{-2}$. The ICL luminosity in this area is 
$L_{\odot,ICL} = 39.2 \times 10^{9} L_{\odot,B}$. From
\citet{binggeli87} the cumulative luminosity from all galaxy types
within 1.3 deg = 80 arcmin radius of the Virgo cluster center is
$L_{\odot,gal} = 58.16 \times 10^{10} L_{\odot,B}$.  Our best estimate
of the fraction of ICL, defined as the un-bound stellar component in
the Virgo cluster core, amounts to $\sim 7$\% of the total light in
Virgo cluster galaxies.

The current fraction is lower than the recent measurement from
\citet{feldmeier04b}, using a very similar technique. This is not
unexpected: we have corrected the surface brightness measurements in
the inner Virgo core fields by subtracting a significant fraction of
PN bound to bright galaxies halos, which is between 70\% and 50\%, of
the total $N_{PN}$ measured in those fields. We think that the current
(conservative) ICL fraction is a robust estimate of the light from
true-unbound stars within 80 arcmin centred on M87.

\subsection{Implications for the formation of the ICL in the Virgo cluster}

We can summarize the current results as follows:
\begin{itemize}
\item The ICL in Virgo is concentrated in the core, with $\mu_{ICL,B}$ in the
range $28.8 - 30 $ mag~arcsec$^{-2}$.
\item In several fields at $R > 80' = 340$ kpc we can set only 
upper limits to the ICL surface brightness. 
\item At $R > 80'$, the ICL signal is confined to isolated fields
(RCN1,LPE) 
\item Diffuse stellar light defined as extended halo plus ICL is
  present in other Virgo sub-structures (central region of sub-clump
  B, M60/M59 sub-group).
\end{itemize}

\begin{figure}
   \centering
   \includegraphics[width=9cm]{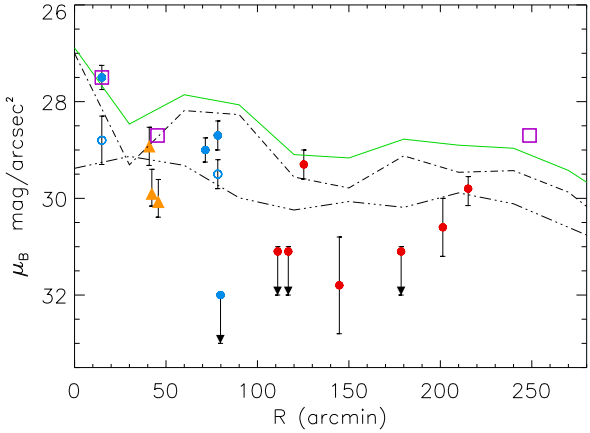}

      \caption{ Surface brightness measurement of diffuse light in the
      Virgo fields (points) compared with the surface brightness
      profile of the Virgo galaxies averaged in annuli (lines); radial
      distances are computed with respect to M87.  The green line
      represents the radial surface brightness profile from light in
      Virgo galaxies from \citet{binggeli87}.  The dotted-dashed and
      double dotted-deshed lines correspond to the surface brightness
      profile associated with giants and dwarf galaxies,
      respectively. The full blue dots show the surface brightness
      measurements in the FCJ, Core, SUBC and LPC fields. The open
      circles at the FCJ, SUBC field position indicates the ICL
      surface brightness computed from the ICPNe not bound to galaxy
      halos.  The triangles represent the surface brightness of the
      ICL based on IC RGB star counts \citep{ferguson98, durrell02,
      williams07}. The full red dots show the surface brightness
      measurements in the RCN1, LPE, SUB2, and SUB3 fields. The arrows
      indicate the upper limits for the measurements in LPRX, LPS and
      LSF.  The magenta open squares indicate the surface brightness
      average values $\mu_B$ at 15, 50 and 240 arcmin computed from
      the measurements listed in Table~5, from \citet{feldmeier04b}
      data; the measurements at $240$ arcmin (F04-2 and F04-6) are
      close to M49, see Figure~\ref{mapvirgo}. Radial distances are
      computed with respect to M87.\label{figure6}}
   \end{figure}

The distribution of the ICL in the core and at larger scales in the
Virgo cluster shows that the ICL is more centrally concentrated than
the galaxies, see Fig.~\ref{figure6}. This result is consistent with
the measurements of \citet{zibetti05} and the predictions of
\citet{murante04, murante07}.

The ICL distribution shows a high degree of substructures both in the
Virgo core and at larger radii. This result is supported by deep
surface brightness photometry \citep{mihos05}, and ICPN number
counts. The presence of sub-structures is predicted by cosmological
simulations of structure formation
\citep{napolitano03,sommerlarsen05,murante07}. The detection of
extended halos and intracluster/intra-group light in Virgo
sub-structures supports a hierarchical formation mechanism for the ICL
as described by \citet{rudick06}: we are observing the ICL component
in sub-structures, which may then become completely unbound as these
sub-structures fall into the densest part of the cluster.

Current data are consistent with most ICL being associated to the
formation history of the brightest cluster galaxies, which explains
the central concentration of the ICL in the core and to
sub-structures. Tidal effects are indeed at work at larger distances
from the center and they can generate some ICL. The ICPNe detected in
RCN1, LPE and SUB2/3 would likely come from such processes, but the
amount of light detected in these fields is small compared with that
sampled close to the big ellipticals, as supported by the
observational results of this work.

%-------------------------------------------------------------
%
%_____________________________________________________________

\section{The Local Mean Luminous density}
From the lack of PN detections in several of our fields in the Virgo
cluster, we can derive a limit on the luminosity density in an assumed
homogeneous distribution of stars in the local volume.
\citet{fukugita98} compiled measurements of the mean luminous density
contributed by high surface brightness galaxies at moderately low
redshift, and adopted $\epsilon_0 = 2 \times 10^{8} h L_{B,\odot}
Mpc^{-3}$.  In what follow, we make the assumption that the mean
luminous density is distributed uniformly in the local volume and
compute the number of PNe associated with this diffuse component in
the surveyed area of the Virgo outer fields and at the [OIII] limiting
magnitude $m_{lim}(5007)=27.0$

Each volume element along the LOS to the Virgo fields contribute the
following number of PNe
\begin{equation}
dN_{PN}(r) = \alpha(r) \times dL(r)
\label{npnr}
\end{equation}
where $L(r)$ is the luminosity associated with the volumetric luminous density 
\begin{equation}
dL(r) = \epsilon_0 \times S \times dr = \frac{\epsilon_0 \Omega}{206265^2}r^2dr 
\end{equation} 
$\epsilon_0 = 2 \times 10^{8} h L_{B,\odot} Mpc^{-3}$ \cite[see
][]{fukugita98}, $\Omega$ is the area in arcsec$^2$, $\alpha(r)$ is
the luminosity-specific planetary nebula number and predicts the
number of PNe associated with the total luminosity $L(r)$ at radial
distance $r$. Its value depends on the relative depth of the on-band
survey ($m_{lim}$) with respects to the faint limit of the PNLF,
e.g. the apparent magnitude of $M^*+8$ at the distance $r$.

The total number of PNe associated with the volumetric density is then
given by the integral of eq.~\ref{npnr}
\begin{equation}
N_{PN,tot}=\int_{0}^{r_{max}} N_{PN}(r) dr
= \int_{o}^{r_{max}}\alpha(r) L_{T}(r) dr =
\label{pntot}
\end{equation}
$$
=\frac{\epsilon_0\Omega}{206265^2} \int_{o}^{r_{max}}\alpha(r)r^2dr
$$ and $r_{max}$ is set by the survey depth, and it is the distance at
which the apparent magnitude of the PNLF bright cut--off $m^*$ is equal
to $m_{lim}$.

{\it The luminosity-specific planetary nebula number $\alpha$ - } The
PN population (N$_{PN}$) associated to a stellar population with
luminosity $L_{T}$ is :
\begin{equation}
N_{PN}=\int_{M^*}^{M^*+8} PNLF(M) dM = 
\label{PNLF}
\end{equation}
$$ =\int_{M^*}^{M^*+8} ke^{0.307M}(1.0-e^{(3*(M^*-M))}) dM = \alpha
L_{T}$$ where $M^{*}=-4.5$ is the absolute magnitude of the bright
cut-off of the PNLF, and $\alpha$ is the luminosity-specific PN
number. $\alpha$ is given by \citet{buzzoni06} for stellar populations
with different ages and metallicities. The maximum theoretical value
of $\alpha$ which gives the {\it largest} PN population for a given
luminosity is $\alpha_{max} = 1 PN \times (1.85 \times 10^6
L_\odot)^{-1}$. From eq.~\ref{npnr} we see that the PNLF extends over
8 magnitudes, but one sees smaller magnitude ranges as distance
increases.  The $\alpha$ parameter must then be corrected using the
double-exponential PNLF analytical formula in eq.~\ref{PNLF} for the
appropriate faint cut-off magnitude set by the the minimum between
$m_{lim}$ and the apparent magnitude corresponding to $M^*+8$ at
distance $r$.

{\it Predicted number of PNe - } We use eq.~\ref{pntot} to compute the
number of PNe we would predict if the \citet{fukugita98} density was
homogeneously distributed.  The number of PNe that we would detect by
surveying 1.75 deg$^2$ (total area covered by the L* fields plus SUB2
and SUB3 fields) down to $m_{lim}=27.0$, is $N_{PN,tot}= 4.6$ with an
expected m(5007) magnitude distribution that peaks at $\sim 26.0$.  We
conclude that we cannot set limits on the homogeneously distributed
luminous density in the local universe, and surveys of several
hundreds square degrees are required to detect a statistical
significant numbers of the brighter PNe.

\section{Conclusions}
We have surveyed several regions of the Virgo cluster located at
different distances from M87, covering a total area of 3.3 square
degrees. These outer fields in the Virgo cluster were imaged through
off- and on-band filters designed for detecting PNe at the distance of
the Virgo cluster. The result of this survey provides an homogeneous
large scale study of the diffuse stellar light in Virgo.

The results from the current study indicate that the diffuse stellar
component is present in the central region of the cluster, within a
radial distance of 80 arcmin from the cluster center. The intracluster
light in the core has a surface brightness of $\mu_{B} = 28.8 - 30$
mag arcsec$^{-2}$, and on average it amounts to $\sim 7\%$ of the
total galaxy light in this region. The distribution of the ICL in the
cluster central region is not homogeneous, with areas where
only upper limits on the number of PNe can be set. This non-uniformity
indicates that the ICL is not relaxed in the core.

At a distances larger than 80 arcmin from M87, the surface brightness
profile of the diffuse light has a sharp cut-off of about 2
magnitudes. At these larger radii the detection of ICL is confined in
single fields, e.g. RCN1, LPE and SUB2/3, or associated with local
substructures, e.g.  sub-clump B, the M60/M59 sub-group. The
observations indicate that the ICL in Virgo is more centrally
concentrated than cluster galaxies, and does not support an
homogeneous elongated distribution of the ICL similar to the late-type
Virgo galaxies \citep{yasuda97, solanes02}.

The concentration of the ICL toward the cluster high density regions
links the formation of the ICL with the formation and evolution of the
most luminous cluster galaxies \citep{rudick06, murante07}.

\begin{acknowledgements}
The authors would like to thank the anonymous referee for the careful
reading of the manuscript and the extensive report. JALA wish to thank
the travel support from ESO Director Discretionary Funds 2008 during
the writing process of this manuscript. This article is based on
observations made with 1) the Isaac Newton Telescope operated on the
island of La Palma by the Isaac Newton Group in the Spanish del Roque
de los Muchachos Observatory; 2) the ESO/MPI 2.2 m telescope at La
Silla (Chile); and 3) the Subaru Telescope, which is operated by the
National Astronomical Observatory of Japan.
\end{acknowledgements}

\bibliographystyle{aa}
\bibliography{ICPNrefs}

\appendix
\section{Catalogs of the point-like emission line objects detected in the
current survey}

\begin{table}[!h]
\caption{Emission-line objects detected in the LPRX field}
\centering
\begin {tabular}{c c c c}
\hline
\hline
  Name & $\alpha (J2000)$ & $\delta (J2000)$ &m(5007) \\
\hline 
RCN1-1  &   12:25:49.69  &  14:22:08.86  &    25.478  \\   
RCN1-2  &   12:27:11.53  &  14:06:36.37  &    25.913  \\  
RCN1-3  &   12:27:04.63  &  14:14:55.15  &    25.915  \\  
RCN1-4  &   12:26:01.43  &  14:19:13.80  &    25.926  \\  
RCN1-5  &   12:26:11.00  &  14:08:49.73  &    25.926  \\  
RCN1-6  &   12:26:30.36  &  14:07:31.63  &    26.194  \\  
RCN1-7  &   12:26:40.40  &  14:08:35.24  &    26.240  \\  
RCN1-8  &   12:27:12.32  &  14:01:27.14  &    26.249  \\  
RCN1-9  &   12:26:30.01  &  14:17:09.80  &    26.295  \\  
RCN1-10 &   12:25:34.38  &  13:58:09.08  &    26.297  \\  
RCN1-11 &   12:26:05.27  &  14:11:14.95  &    26.311  \\  
RCN1-12 &   12:25:35.01  &  13:59:42.98  &    26.312  \\  
RCN1-13 &   12:27:13.55  &  14:17:48.79  &    26.343  \\  
RCN1-14 &   12:26:59.59  &  14:09:56.40  &    26.348  \\  
RCN1-15 &   12:27:08.51  &  14:08:29.50  &    26.355  \\  
RCN1-16 &   12:26:59.25  &  13:53:42.86  &    26.359  \\  
RCN1-17 &   12:26:28.77  &  14:08:34.98  &    26.394  \\  
RCN1-18 &   12:26:33.17  &  14:16:20.35  &    26.489  \\  
RCN1-19 &   12:26:28.15  &  14:08:04.62  &    26.490  \\  
RCN1-20 &   12:25:55.82  &  14:09:08.78  &    26.572  \\  
RCN1-21 &   12:27:04.90  &  14:10:43.36  &    26.602  \\  
RCN1-22 &   12:25:54.25  &  13:55:17.77  &    26.657  \\  
RCN1-23 &   12:27:15.57  &  14:23:12.65  &    26.674  \\  
RCN1-24 &   12:27:07.64  &  14:16:10.91  &    26.681  \\  
RCN1-25 &   12:25:27.23  &  14:11:33.21  &    26.682  \\  
RCN1-26 &   12:27:07.01  &  13:59:56.88  &    26.686  \\  
RCN1-27 &   12:26:34.64  &  14:11:55.75  &    26.689  \\  
RCN1-28 &   12:26:09.25  &  13:57:22.98  &    26.688  \\
\hline
\end {tabular} 
\label{Tab:cand_virgo_RCN1}
\end{table}

\begin{table}[!h]
\caption{Emission-line objects detected in the LPRX field}
\centering
\begin {tabular}{c c c c}
\hline
\hline
  Name & $\alpha (J2000)$ & $\delta (J2000)$ &m(5007) \\
\hline 
LPRX-1  &12:27:20.377 & 9:17:37.14 &  26.03  \\
LPRX-2  &12:26:32.996 & 9:23:48.76 &  26.42  \\
LPRX-3  &12:28:11.142 & 9:25:14.66 &  26.47  \\  
LPRX-4  &12:27:25.608 & 9:16:14.40 &  26.53  \\
LPRX-5  &12:27:48.012 & 9:14:26.81 &  26.78  \\
LPRX-6  &12:26:42.855 & 9:16:12.58 &  26.94  \\
LPRX-7  &12:27:25.148 & 9:12:53.31 &  27.19  \\
LPRX-8  &12:27:31.843 & 9:37:10.15 &  27.33  \\     
LPRX-9  &12:26:40.641 & 9:28:57.08 &  27.36  \\ 
\hline
\end {tabular} 
\label{Tab:cand_virgo_lprxx}
\end{table}

\begin{table}[!h] 
\caption{Emission-line objects detected in the LPS field}
\centering
\begin{tabular}{c c c c}
\hline \hline
Name&  $\alpha (J2000)$ & $\delta (J2000)$ & m(5007) \\
\hline 
LPS-1 &12:27:25.204 & 10:51:38.34 & 	  26.21 \\
LPS-2 &12:26:38.216 & 10:35:49.34 & 	  26.36  \\
LPS-3 &12:27:34.998 & 10:26:46.14 & 	  26.38 \\
LPS-4 &12:27:10.849 & 10:43:44.31 & 	  26.75  \\
LPS-5 &12:27:33.526 & 10:44:30.01 & 	  26.98  \\
LPS-6 &12:25:59.211 & 10:44:33.17 & 	  27.01  \\
LPS-7 &12:27:38.171 & 10:53:50.34  &	  27.10  \\
\hline
\end {tabular} 
\label{Tab:cand_virgo_lps}
\end {table}

\begin{table}[!h]   
\caption{Emission-line objects in the LPE field}
\centering
\begin{tabular}{c c c c}
\hline
\hline
Name&  $\alpha (J2000)$ & $\delta (J2000)$ & m(5007)  \\
\hline 
LPE-1 &  12:17:58.471 &  13:40:07.24 & 26.08\\
LPE-2 &  12:17:32.446 &  13:48:31.87 & 26.15\\
LPE-3 &  12:17:39.261 &  13:32:27.24 & 26.19\\
LPE-4 &  12:18:04.059 &  13:43:38.15 & 26.19\\
LPE-5 &  12:17:59.871 &  13:43:18.95 & 26.40\\
LPE-6 &  12:17:17.932 &  13:28:25.66 & 26.43\\
LPE-7 &  12:16:04.795 &  13:40:59.73 & 26.44\\
LPE-8 &  12:16:36.738 &  13:45:45.07 & 26.52\\
LPE-9 &  12:16:02.328 &  13:31:03.16 & 26.71\\
LPE-10 &  12:17:01.922 &  13:56:48.82 & 26.8\\
LPE-11 &  12:18:04.406 &  13:43:05.79 & 26.8\\
LPE-12 &  12:16:54.828 &  13:58:22.60 & 26.9\\
LPE-13 &  12:18:14.529 &  13:39:13.96 & 27.0\\
LPE-14 &  12:18:07.388 &  13:52:02.52 & 27.0\\
LPE-15 &  12:16:53.685 &  13:29:48.63 & 27.1\\
LPE-16 &  12:16:48.095 &  13:52:14.56 & 27.2\\
LPE-17 &  12:16:59.375 &  13:44:38.69 & 27.3\\
LPE-18 &  12:16:41.977 &  13:34:25.83 & 27.3\\
LPE-19 &  12:18:09.217 &  13:50:04.31 & 27.4\\
LPE-20 &  12:18:11.198 &  13:57:59.73 & 27.6\\
\hline
\end {tabular} 
\label{Tab:cand_virgo_lpe}
\end {table}

\begin{table}[!h]
\caption{Emission-line objects detected in the LSF field}
\centering
\begin{tabular}{c c c c}
\hline
\hline
Name&  $\alpha (J2000)$ & $\delta (J2000)$ & m(5007)   \\
\hline 
    LSF-1 &  12:37:42.215 &12:15:43.10 &  25.99     \\
    LSF-2 &  12:39:13.235 &12:14:38.27 &  26.83     \\
    LSF-3 &  12:38:07.374 &12:07:34.85 &  27.08     \\
    LSF-4 &  12:39:13.669 &12:01:46.78 &  27.17    \\
    LSF-5 &  12:37:42.473 &12:11:20.97 &  27.35    \\
    LSF-6 &  12:39:20.662 &12:16:37.00 &  27.36    \\
    LSF-7 &  12:39:09.730 &12:10:50.76 &  27.37    \\
    LSF-8 &  12:39:13.045 &12:11:06.79 &  27.66    \\
    LSF-9 &  12:37:46.282 &11:55:14.22 &  27.70    \\
    LSF-10&  12:38:47.034 &12:18:00.90 &  27.70    \\
    LSF-11&  12:37:28.577 &12:17:48.96 &  27.80     \\
    LSF-12&  12:37:46.895 &12:15:24.38 &  27.91    \\
    LSF-13&  12:37:38.274 &12:20:31.31 &  27.93    \\
    LSF-14&  12:39:15.965 &11:55:43.51 &  27.93    \\
    LSF-15&  12:37:30.846 &12:19:05.18 &  27.97 	 \\
    LSF-16&  12:39:27.344 &12:20:25.23 &  27.99    \\
    LSF-17&  12:39:20.736 &12:05:44.44 &  28.02    \\
    LSF-18&  12:37:29.078 &12:22:37.17 &  28.03    \\
    LSF-19&  12:38:05.800 &12:01:34.96 &  28.03    \\
    LSF-20&  12:38:41.080 &11:57:07.98 &  28.04    \\
    LSF-21&  12:38:00.317 &12:21:35.34 &  28.06    \\
    LSF-22&  12:38:37.500 &12:05:56.12 &  28.09    \\
    LSF-23&  12:39:23.081 &12:12:33.49 &  28.15    \\
\hline
\end {tabular} 
\label{Tab:cand_virgo_lsf}
\end {table}

\begin{table*}[!h]   
\caption{Emission-line objects detected in the SUB2 field. Objects
  with (*) are ICPNe}
\centering
\begin{tabular}{c c c c c c}
\hline
\hline
Name&  $\alpha (J2000)$ & $\delta (J2000)$ & m(5007) & m$_{H\alpha}$ & m$_{(V+R)}$\\
\hline 
SUB2-1&   12:23:22.524 &14:03:21.68    & 26.30  &--- & ---  \\
SUB2-2&   12:24:23.113 &14:20:50.64    & 27.01  &--- & 30.22  \\
SUB2-3&   12:24:45.272 &14:18:52.58    & 27.18  &--- & 29.91  \\
SUB2-4&   12:24:35.834 &14:05:36.93    & 27.19  &--- & 29.26  \\
SUB2-5&   12:24:45.103 &14:16:12.88    & 27.34  &--- & 27.56  \\
SUB2-6&   12:24:46.116 &14:13:04.52    & 27.46  &--- & ---  \\
SUB2-7*&   12:24:15.861 &14:04:17.51   & 27.54  &25.81 & 28.33  \\
SUB2-8  & 12:25:01.022 & 14:02:10.35   & 27.58 & ---  & 28.53\\
SUB2-9&  12:23:39.559 &14:18:15.09     & 27.59  &--- & 31.73  \\
SUB2-10&  12:24:35.553 &13:55:21.40    & 27.75  &--- & ---  \\
SUB2-11&  12:23:59.558 &14:13:02.89    & 27.85  &--- & ---  \\
SUB2-12&  12:23:04.010 &14:06:12.86    & 27.88  &--- & ---  \\
SUB2-13&  12:23:25.975 &14:05:48.30    & 27.90  &--- & ---  \\
\hline
\end{tabular} 
\label{Tab:cand_virgo_sub2_cm}
\end{table*}

\begin{table*}[!h]   
\caption{Emission-line objects detected in the SUB3 field. Objects
  with (*) are ICPNe}
\centering
\begin{tabular}{c c c c c c}
\hline
\hline
Name&  $\alpha (J2000)$ & $\delta (J2000)$ & m(5007) & m$_{H\alpha}$ & m$_{(V+R)}$\\
\hline 
SUB3-1 &  12:24:21.128 &15:32:10.67  &  24.57  &---&  --- \\
SUB3-2 &  12:25:02.814 &15:16:42.17  &  25.75  &---&  --- \\
SUB3-3 &  12:25:05.471 &15:16:30.21  &  26.13  &---&  --- \\
SUB3-4 &  12:25:05.812 &15:16:28.84  &  26.25  &---&  28.10 \\
SUB3-5* &  12:24:26.934 &15:15:43.10  &  26.31 &25.46&  27.30 \\
SUB3-6* &  12:25:05.369 & 15:23:04.51 &  26.46 &25.46  &27.66 \\
SUB3-7* &  12:23:32.742 & 15:16:58.09 &  26.49 &---  &27.06 \\
SUB3-8 &  12:25:15.651 &15:27:13.45  &  26.57  &---&  26.86 \\
SUB3-9 &  12:25:00.771 &15:15:52.94  &  26.99  &---&  26.72 \\
SUB3-10&  12:24:35.389 &15:08:58.35  &  27.52  &---&  28.72 \\
SUB3-11&  12:23:40.561 &15:09:28.04  &  27.53  &---&  --- \\
\hline
\end{tabular} 
\label{Tab:cand_virgo_sub3_cm}
\end{table*}

\end{document}